\documentstyle[12pt]{article}
\newcommand{\pgv}{\mbox{\boldmath $\pi $}}
\newcommand{\ngv}{\mbox{\boldmath $\nabla $}}
\newcommand{\p}{\partial}
\newcommand{\pv}{\mbox{\bf p}}
\newcommand{\xv}{\mbox{\bf x}}
\newcommand{\vp}{\varphi}
\begin{document}
\title{\bf
On the Canonical Treatment of  Lagrangian Constraints
}
\author{B.M.\ Barbashov \bigskip \\
{\it
Joint Institute for Nuclear Research} \\
{\it Dubna, 141980 Russia
}}
\date{}
\maketitle
\begin{abstract}
The canonical treatment of dynamic systems with manifest
Lagrangian constraints proposed by Berezin is applied to concrete
examples:  a special Lagrangian linear in velocities, relativistic
particles in proper time gauge, a relativistic string in
orthonormal gauge, and the Maxwell field in the Lorentz gauge,
\end{abstract}

The conventional canonical treatment of constrained
systems~\cite{Dirac}  deals with the constraints which follow only
from the intial singular Lagrangian. However, there are problems
where the Lagrange constraints are introduced `by hand' in addition
to the initial Lagrangian or when from the very beginning of the
Hamiltonization procedure, some of the constraints that follow from
the Lagrange function, are taken into account manifestly.
%
For example, the Lorentz
gauge in electrodynamics cannot be canonically implemented~\cite{Yaffe}. The
purpose of this note is to show that such "noncanonical" constraints
can be implemented by the Berezin algorithm~\cite{Berezin}. The
algorithm provides a unified consideration of the singular
Lagrangians and nonsingular ones with constraints that depend on
velocities and time:
\begin{equation}
\label{eq-1}
 \varphi_a (q,\dot
q,t)=0,\;\; q=(q_1,q_2, \ldots , q_n), \; a=1,2, \ldots  , m, \;
m\leq n {.}
\end{equation}

Let us consider the Lagrangian $L(q, \dot q, t)$ and the set of
the Lagrangian constraints (\ref{eq-1}).
The relevant extended (generalized) Lagrangian reads
\begin{equation}
\label{eq-2}
{\cal L} (q,\dot q ,t)=L(q, \dot q,t)+\sum_{a=1}^{m}\lambda _a(t)\varphi _a
(q,\dot q,t)
\end{equation}
where $\lambda_a$ are the Lagrange multipliers. All the constraints to be
considered depend explicitly on velocities $\dot q_i$. When among them, there
exists the equation $\varphi (q,t)=0$, we replace it, after differentiating
with respect
to time, by the equivalent equation
\begin{equation}
\label{eq-1a}
\sum_{i=1}^{n}\frac{\partial \varphi }{\partial q_i}\dot q_i +\frac{\partial
\varphi}{\partial t }=0{.}
\end{equation}
Now, let us introduce the extended momenta for the Lagrangian function (2)
\begin{equation}
\label{eq-3}
\tilde p_i=\frac{\partial {\cal L}}{\p \dot q_i}=\frac{\p L}{\p \dot q_i}+\sum_{a=1}^{m}
\lambda_a\frac{\p \vp_a}{\p \dot q_i}, \quad  i=1, \ldots , n{.}
\end{equation}
Berezin~\cite{Berezin} has assumed that the velocities $\dot q_i$ and the
Lagrange
multipliers $\lambda_a(t)$ can be expressed uniquely in terms of $q_i$ and
$ \tilde p_i$ by resolving the constraints (1) together with
Eqs.\ (\ref{eq-3}). In this case, the variational problem is said to be a
nondegenerate (nonsingular) one. On the contrary, the requirement of the
initial Lagrangian being nonsingular
\begin{equation}
\label{eq-4}
\det \left | \left |
\frac{\p ^2 L}{\p \dot q_i \p \dot q_j}
\right | \right |
  \not = 0
\end{equation}
becomes superfluous. In the following, only the dynamic systems that satisfy
the Berezin assumption will be considered. The method does not lead to the
reduction of degrees of freedom of the systems in the phase space. However,
the transition to the canonical momenta $p$, corresponing to
the initial Lagrangian $L$ takes place if
\begin{equation}
\label{eq-5}
\left .
\lambda _a(\tilde p,q,t)\right |_{\tilde p= p} =0\,{.}
\end{equation}
 It will lead to
the primary Hamiltonian constraints in this approach. As an
illustration, we apply the Berezin method to a number of constrained
Lagrangian systems.

{\bf 1.} The Lagrangian linear in velocities~\cite{Berezin,NB,FJ}
\begin{equation}
\label{eq-1.1}
L=\sum_{i=1}^{n}f_i(q) \dot q_i- V(q), \quad q=(q_1,q_2, \ldots , q_n){.}
\end{equation}
Since the Lagrangian (\ref{eq-1.1}) is singular, and all the equations of motion
\begin{equation}
\label{eq-1.2}
\sum_{j=1}^{n}f_{ij}\dot q_j +\frac{\p V}{\p q_i}=0,\;\;
f_{ij}=\frac{\p f_i}{\p q_j}-\frac{\p f_j}{\p q_i},\;\;
\det  \parallel f_{ij}\parallel \not=0
\end{equation}
become first-order equations, the extended Lagrangian acquires the form
\begin{equation}
\label{eq-1.3}
{\cal L}= \sum_{i=1}^{n}f_i(q)\dot q_i-V(q)+
\sum_{i,j=1}^{n} \lambda_i\left (f_{ij}\dot q_j +\frac{\p V}{\p \dot q_i}
\right )
\end{equation}
and the extended momenta read
\begin{equation}
\label{eq-1.4}
\tilde p_i=\frac{\p {\cal L}}{\p q_i}= f_i(q)+\sum_{j=1}^{n}\lambda _j
f_{ji}(q){.}
\end{equation}
It is possible to resolve Eqs.\  (\ref{eq-1.2}) with respect
to $\dot q_i$, because
there exists the inverse matrix $f^{-1}_{ij}$ such that
\begin{equation}
\label{eq-1.5}
\dot q_i =- \sum_{j=1}^{n}f^{-1}_{ij}\frac{\p V}{\p q_j}{.}
\end{equation}
Also, resolution of (\ref{eq-1.4}) with respect to $\lambda_i$ gives us
\begin{equation}
\label{eq-1.6}
\lambda_{_i}=\sum_{j=1}^{n}f^{-1}_{ij}(\tilde p_j-f_j){.}
\end{equation}
Taking into account that ${\cal L}$ in
(\ref{eq-1.3}) on the surface of constraints has
the form
$$
{\cal L}=\sum_{i,j=1}^{n}f_if_{ij}\frac{\p V}{\p q_j}-V(q).
$$
we find that
\begin{equation}
\label{eq-1.7}
{\cal H}=\sum_{i=1}^{n}\tilde p_i\dot q_i-{\cal L}=
\sum_{i,j=1}^{n}(f_i -\tilde p_i)f^{-1}_{ij}\frac{\p V}{\p q_j}+V(q){.}
\end{equation}
Going over to the canonical momenta $p$, from (\ref{eq-5}) and (\ref{eq-1.6}),
we obtain the
primary Hamiltonian constraints (invariant relations)
\begin{equation}
\label{eq-1.8}
\lambda_i |_{\tilde p=p} =0 \Longrightarrow p_i=f_i(q){.}
\end{equation}
The kinetic term in (\ref{eq-1.7}) is linear in $\tilde p$,
thus,  ${\cal H}$ is singular.
Therefore,
again there is no Legendre transformation from ${\cal H}$ to ${\cal L}$
because the relations
\begin{equation}
\label{eq-1.9}
\dot q_i=\frac{\p {\cal H}}{\p p_i}=-\sum_{j=1}^{n}f_{ij}\frac{\p V}{\p q_j}
\end{equation}
do not contain $p$.
However, with the help of
the Berezin algorithm, the system (\ref{eq-1.7}), (\ref{eq-1.8})
can be transformed into the
initial Lagrangian system. Indeed, we derive the extended Hamiltonian in the
form
$$
{\cal H}_{ext}={\cal H}+\sum_{i=1}^{n}\mu_i(p_i-f_i(q))
\Longrightarrow\tilde{\dot q}=
-\sum_{j=1}^{n}f^{-1}_{ij}\frac{\p V}{\p q_j}+\mu_i{.}
$$
Thus, we arrive at the system of equations
\begin{equation}
\label{eq-1.10}
\mu_i =\tilde {\dot q_i}+\sum_{j=1}^{n}f^{-1}_{ij}\frac{\p V}{\p q_j}, \quad p_i
=f_i(q)
\end{equation}
and can
construct the Lagrangian
$$
L=\sum_{i=1}^{n}p_i\tilde{\dot q_i}-{\cal H}_{ext}=
\sum_{i=1}^{n}\tilde {\dot q}_if_i(q)-V(q)\,{.}
$$
Going over to the generalized velocities
$\tilde{\dot q}\to \dot q$ via the equation
$\mu_i |_{\tilde{\dot q}=q}=0$, from (\ref{eq-1.7}) we once again
obtain the Lagrangian
constraints~(\ref{eq-1.5}).
{\bf 2.} Relativistic point particle $L=-m \int \sqrt{\dot x^2(\tau )}
d\tau$~\cite{OT}.

If the parameter $\tau$ is chosen as the proper time, the Lagrangian
constraint is $\dot x^2=c^2=\mbox{constant}$. The density of the extended Lagrangian for this system
takes the form
\begin{equation}
\label{eq-2.1}
{\cal L}=-m\sqrt{\dot x^2(\tau )}-\lambda \frac{m}{2}[\dot x^2(\tau )-c^2],
\end{equation}
\begin{equation}
\label{eq-2.2}
\tilde p_{\mu}=-\frac{\p {\cal L}}{\p \dot x^\mu}=m\left [\frac{\dot x_
\mu}{\sqrt{\dot x^2}}+\lambda \dot x_\mu
\right ].
\end{equation}
As before, we use equations (\ref{eq-2.2}) and the above
constraint to find $\lambda$
and $\dot x_{\mu}$
\begin{equation}
\label{eq-2.3}
\lambda=\frac{\sqrt{\tilde p^2}-m}{cm}, \;\; \dot x_\mu =c\frac{\tilde p_\mu}
{\sqrt{\tilde p^2}}
{.}
\end{equation}
As a result, we arrive at the Hamiltonian in the following form (taking into
account that on the constraint shell, ${\cal L} = -mc$):
\begin{equation}
\label{eq-2.4}
{\cal H}=-\tilde p^\mu \dot x_\mu -{\cal L}=c(m-\sqrt{\tilde p^2}){.}
\end{equation}
Turning to the canonical momenta $\lambda |_{\tilde p=p}=0$, from (\ref{eq-2.3}), we get the Hamiltonian
constraint
\begin{equation}
\label{eq-2.5}
\sqrt{p^2}=m{.}
\end{equation}
The Hamiltonian equations for (\ref{eq-2.4}) coincide  with the Lagrangian
ones:
$$
\dot x_\mu =-\frac{\p {\cal H}}{\p p^\mu}=c \frac{p_\mu}{\sqrt{p^2}}=
c\frac{p_\mu}{m};\;\;
\dot p_\mu =\frac{\p {\cal H}}{\p x^\mu}=0{.}
$$
The obtained Hamiltonian (\ref{eq-2.4}), as well as the Lagrangian, is singular, but
applying the same algorithm, it is possible to restore the initial Lagrangian
system:
$$
{\cal H}=c(m-\sqrt{p^2})+\mu (m-\sqrt{p^2})\;\;
\Longrightarrow
{\tilde{\dot x}}_\mu=(c+\mu)\frac{p_\mu}{\sqrt{p^2}}{.}
$$
Adding the constraint (\ref{eq-2.5}) to this system, we obtain
$$
\mu=\sqrt{{\tilde{\dot x}}^2}-c, \quad p_\mu =m\frac{{\tilde{\dot x}}_\mu}
{\sqrt{{\tilde{\dot x}}^2}}{.}
$$
Finally,
$$
{\cal L}=-p_\mu{\tilde {\dot x}}^\mu -{\cal H}=-m\sqrt{{\tilde
{\dot x}}^2},\;\;
\mu|_{\tilde{\dot x}=\dot x}=0\;
\Longrightarrow \sqrt {\dot x^2}=c{.}
$$

\noindent
{\bf 3.} Relativistic particle with the gauge $x^0={\cal P}\tau /m$.

Differentiating this gauge with respect to time $\dot x^0={\cal P}/m$
and substituting it
into the extended Lagrangian, we obtain
\begin{equation}
\label{eq-3.1}
{\cal L}=-m\sqrt{
(\dot x^0)^2- {\dot {\mbox{\bf x}}}^2
}-\lambda m(\dot x^0-{\cal P}/m){,}
\end{equation}
\begin{equation}
\label{eq-3.2}
\tilde p_0=-\frac{\p {\cal L}}{\p \dot x^0}=m \left (
\frac{\dot x_0}{\sqrt{\dot x^2}}+\lambda
\right ); \quad \mbox{\bf p}=\frac{\p {\cal L}}{\p \dot {\mbox{\bf x}}}=
m\frac{\dot {\mbox{\bf x}}}{\sqrt{\dot x^2}}{.}
\end{equation}
Applying once more the condition $\dot x^0={\cal P}/m$, we have
$$
\tilde p_0=\frac{{\cal P}}
{\sqrt{
({\cal P}/m)^2-{\dot{\mbox{\bf x}}}^2
}}
+\lambda m,\quad \mbox{\bf p}=\frac{m\dot{\mbox {\bf x}}}
{\sqrt{
({\cal P}/m)^2-{\dot{\mbox{\bf x}}}^2
}}{.}
$$
Hence, it follows that
\begin{equation}
\label{eq-3.3}
\lambda =\frac{\tilde p_0-\sqrt{{\mbox{\bf p}^2}+m^2}}{m},\quad
\dot{\mbox{\bf x}}=\frac{{\cal P}}{m}\frac{\mbox{\bf p}}
{\sqrt{{\mbox{\bf p}^2}+m^2}}{.}
\end{equation}
The Hamiltonian reads
\begin{equation}
\label{eq-3.4}
{\cal H}=-p_0\dot x^0+\pv \dot{\xv}-{\cal L}=
\frac{{\cal P}}{m}(\sqrt{{\mbox{\bf p}^2}+m^2}-\tilde p_0){.}
\end{equation}
Going over to the canonical momentum $p_0$ by means of $\lambda
 |_{\tilde p_0=p_0}=0$, we derive the
Hamiltonian constraint
$p_0=\sqrt{\pv^2+m^2}.$
Then, the Hamiltonian equations are as follows:
$$
 {\dot{x}}^0=-\frac{\p {\cal H}}{\p p_0}=\frac{{\cal P}}{m};\;\;
{\dot {\mbox {\bf x}}}=\frac{\p {\cal H}}{\p p^0}=\frac{{\cal P}}{m}
\frac{\pv}{\sqrt{\pv^2+m^2}}{,}
$$
$$
\dot p_0=\frac{\p {\cal H}}{\p x^0}=0,\quad \dot{\pv }=-\frac{\p {\cal H}}
{\p \xv}=0{.}
$$
{\bf 4.} Relativistic string in the orthonormal gauge~\cite{BN}:
\begin{equation}
\label{eq-4.1}
L=-\gamma \sqrt{(\dot x x')^2-\dot x^2 x'^2},\quad \dot x^2+x'^2=0, \quad (\dot x
x')=0{.}
\end{equation}
The extended Lagrangian  in this case  reads
$$
{\cal L}= -\gamma \sqrt{(\dot x x')^2-\dot x^2 x'^2}-\frac{\lambda_1}{2}(\dot x^2
+x'^2)-\lambda _2(\dot x x'){,}
$$
and taking account of the constraints we get for extended momenta
\begin{equation}
\label{eq-4.2}
\tilde p_\mu= -\frac{\p {\cal L}}{\p \dot x^\mu}= (\gamma + \lambda_1)\dot x_\mu
+\lambda _2 x'_\mu {.}
\end{equation}
Projecting this onto $x_\mu'$ and using the constraints, we find that $(\tilde p x')
=\lambda_2 x'^2$,
and then,
\begin{equation}
\label{eq-4.3}
\lambda_2=\frac{(\tilde p x')}{x'^2}{.}
\end{equation}
Squaring (\ref{eq-4.2}), we obtain
\begin{equation}
\label{eq-4.4}
\tilde p^2=(\gamma+\lambda_1)^2(-x'^2)+\frac{(\tilde p x')^2}{x'^2}\;
\Longrightarrow \lambda_1+\gamma =\frac{\sqrt{(\tilde p x')^2-\tilde p^2 x'^2}}
{-x'^2}{.}
\end{equation}
Given $\lambda_1$ and $\lambda_2$, we can express $\dot x_{\mu}$ in terms of
$x'_\mu $ and $\tilde p_\mu$ as follows:
$$
\dot x_\mu =\frac{(\tilde p x')x'_\mu-x'^2\tilde p_\mu}{\sqrt{(\tilde p x')^2
-\tilde p^2 x'^2}}{,}
$$
which satisfies the constraints identically. As a result, the Hamiltonian for
the string assumes the form
\begin{equation}
\label{eq-4.5}
{\cal H}= -\tilde p^\mu \dot x_\mu-{\cal L} =-\sqrt{(\tilde p x')^2
-\tilde p^2 x'^2} -\gamma\, x'^2{.}
\end{equation}
Going over to the canonical momenta $p_{\mu}$ according to formula (\ref{eq-5}), we
arrive at the Hamiltonian constraints
\begin{equation}
\label{eq-4.6}
\lambda_i\_{\tilde p=p}=0\; \Longrightarrow
(p x')=0,\;\; p^2+\gamma^2x'^2=0{.}
\end{equation}
On the surface of these Hamiltonian constraints, ${\cal H} = 0$, the
canonical equations are as follows:
$$\dot x_\mu =-\frac{\p {\cal H}}{\p p^\mu}=
\frac{(px')x'_\mu-x'^2p_\mu}{\sqrt{(\tilde p x')^2-\tilde p^2 x'^2}}{,}
$$
$$
\dot p_\mu=-\frac{\p }{\p \sigma }\left (
\frac{\p {\cal H}}{\p x'^\mu}
\right )=\frac{\p}{\p \sigma}\left [
\frac{(px')p_\mu -p^2x'_\mu}{\sqrt{(\tilde p x')^2-\tilde p^2 x'^2}}+
2 \gamma x'_\mu
\right ]{.}
$$
Taking account of constraints (\ref{eq-4.6}), we can rewrite
these equations in the form
$$
\dot x_\mu=\frac{1}{\gamma }p_\mu, \quad \dot p_\mu=\gamma
x''_\mu \; \Longrightarrow \ddot x_\mu -x''_\mu =0{.}
$$
The Hamiltonian (\ref{eq-4.5}) is singular
$
\det \left | \left |
\frac{\p ^2 {\cal H}}{\p  p^\mu \p p^\nu }
\right | \right |
   = 0
$. But using the Berezin algorithm, we
can pass to the initial Lagrangian $L$ and constraints (\ref{eq-4.1}). Indeed,
the extended
Hamiltonian is of the form
$$
{\cal H}_{ext}= -\sqrt{(p x')^2- p^2 x'^2} -\gamma x'^2 -
\frac{\mu _1}{2 \gamma}(p^2+ \gamma^2 x'^2) -\mu_2 (px'){,}
$$
from which and the subsidiary conditions we find
$$
\tilde {\dot x}_\mu =\frac{\p {\cal H}_{ext}}{\p p^\mu}= \gamma^{-1}(1+\mu_1)p_\mu+\mu_2
x'_\mu,\Longrightarrow
(\tilde {\dot x} x')=\mu_2 x'^2{,}
$$
\begin{equation}
\label{eq-4.7}
{\tilde{ \dot x}}^2=-(1+\mu_1)^2x'^2+\frac{(\tilde {\dot x}x')^2}{x'^2}\;
\Longrightarrow 1+\mu_1=
\frac{\sqrt{(x'\tilde {\dot x})^2-x'^2{\tilde {\dot x}}^2}}{-x'^2}
\end{equation}
and therefore, from (\ref{eq-4.7}), we get
$$
p_\mu =\gamma \frac{(x'\tilde {\dot x})x'_\mu-x'^2\tilde {\dot x}^2}
{\sqrt{(x'\tilde {\dot x})^2-x'^2{\tilde {\dot x}}^2}}{.}
$$
Finally, we obtain
$$
L=-p_\mu {\tilde {\dot x}}^\mu- {\cal H}_{ext}=
\gamma \sqrt{(x'\tilde{\dot x})^2-x'^2{\tilde {\dot x}}^2}
$$
$$
\mu_i|_{\tilde {\dot x}=\dot x}=0 \Longrightarrow (\dot x x')=0,\;\;
\dot x^2+x'^2=0.
$$
\noindent
{\bf 5.} Electromagnetic field with the Lorentz gauge $\p_\mu A^\mu=0$~\cite{Wentzel}.

The extended Lagrangian with an external source $j^{\mu}$ is of the form
$$
{\cal L}=-\frac{1}{4}F_{\mu \nu}F^{\mu \nu}-j_\mu A^\mu-\lambda(\p _\mu
A^\mu),
$$
$$
F^{\mu \nu}=\p^\mu A^\nu-\p^\nu A^\mu, \quad \p_\mu j^\mu =0{.}
$$
This gives the time component of the extended momenta $\tilde \pi_0$ as follows
\begin{equation}
\label{eq-5.1}
\tilde \pi_0=-\frac{\p {\cal L}}{\p \dot A^0}=\lambda{.}
\end{equation}
For the space component $\pgv$, we derive the canonical expression
\begin{equation}
\label{eq-5.2}
 \tilde{\pgv}=\frac{\p {\cal L}}{\p \dot {{\bf A}}}=\dot{\bf A}+{\bf \ngv}  A^0{.}
\end{equation}
According to the Berezin algorithm, adding the Lorentz gauge
\begin{equation}
\label{eq-5.3}
\dot A^0=-(\ngv {\bf A})
\end{equation}
to these equations, resolving the velocities $\dot A^{\mu}$ and the
multiplier  $\lambda$ in terms of $A^{\mu}$ and $\pi ^\mu$, we obtain
\begin{equation}
\label{eq-5.4}
\lambda={\tilde \pi}^0,\;\; {\bf A} =\pgv -\ngv A^0,\;\; \dot A^0=-(\ngv {\bf A}){.}
\end{equation}
Now, one can construct the Hamiltonian
$$
{\cal H}=\tilde \pi^0 (\ngv {\bf A}) +\frac{1}{2}(\pgv ^2 +(\mbox {rot}
\,{\bf A})^2)-(\pgv \ngv \,A^0)+j_\mu A^\mu {.}
$$
It gives the canonical equations
$$
\dot A^0=-\frac{\p {\cal H}}{\p \tilde \pi _0}=- (\ngv {\bf A}),\quad
\dot {\bf A}=\frac{\p{\cal H}}{\p \pgv}=\pgv -\ngv A^0
$$
coinciding with (\ref{eq-5.2}) and (\ref{eq-5.3}).
Also, for the momenta, we get
$$
\tilde {\dot \pi }_0=\frac{\p {\cal H}}{\p A^0}-
\sum_{j=1}^{3}\frac{\p }{\p x_j}\frac{\p {\cal H}}{\p (\p A^0/\p x_j)}=
j_0+(\ngv \pgv),
$$
\begin{equation}
\label{eq-37}
\dot{ \pgv} =-\frac{\p {\cal H}}{\p {\bf A}}+\sum_{j=1}^{3}\frac{\p}{\p x_j}
\frac{\p {\cal H}}{\p (\p {\bf A}/\p x_j)}={\bf j}+ \ngv \tilde {\pi }^0-
\mbox{rot}\, \mbox{rot}\, {\bf A}{.}
\end{equation}
And for the components of ${\pi}^{\mu}$ and $A^{\mu}$, we have
$$
\left .\ddot A^0 =\Delta A^0 +j^0 -\dot{\tilde \pi}^0 \atop
\ddot {\bf A}=\Delta {\bf A}+{\bf j}+
\ngv \tilde \pi^0\right \}     \longrightarrow
\Box A^\mu =j^\mu -\p ^\mu \tilde {\pi ^0}{;}
$$
\begin{equation}
\label{eq-38}
\Box \pgv =\dot {\bf j} +\ngv j^0;\quad \Box \tilde \pi^0 =0{.}
\end{equation}
After the transition $\lambda |_{{\tilde \pi}^0=\pi^0}=0$, from (\ref{eq-5.4}), we
obtain $\pi^0=0$, and all the equations (\ref{eq-37}) and (\ref{eq-38})
are cast into the correct equations of electrodynamics in the Lorentz gauge.

To conclude, we note that contrary to the Dirac approach, the suggested
algorithm allows unique construction of the Hamilton formalism for
constrained Lagrangian systems with constraints that depend on velocities
and, in the general case, do not depend on the Lagrangian form.

{\bf Acknowledgments.} A am grateful to Professor L.D.~Fadeev and Dr.\
 V.V.~Nesterenko for critical discussions of this work.

\end{document}